\documentclass[useAMS]{mn2e}
\usepackage{amssymb,amsmath,psfig,times}
\usepackage{graphicx}
\def\gsim{ \lower .75ex \hbox{$\sim$} \llap{\raise .27ex \hbox{$>$}} }
\def\lsim{ \lower .75ex\hbox{$\sim$} \llap{\raise .27ex \hbox{$<$}} }

\def\beq{\begin{equation}}
\def\eeq{\end{equation}}

\def\hete{{\it Hete--II}}
\def\sw{{\it Swift}}

\def\cgro{{\it CGRO}}

\def\ep{$E_{\rm p}$}

\def\eiso{$E_{\gamma,\rm iso}$}
\def\ama{$E_{\rm p}-E_{\rm iso}$}

\def\ghi{$E_{\rm p}-E_{\gamma}$}

\def\th{$\theta_{\rm j}$}

\def\thv{$\theta_{\rm v}$}
\def\tjet{$t_{\rm break}$}

\def\egamma{$E_{\gamma}$}

\def\G{$\Gamma_{0}$}

\title[GRB jet structure]
{Luminosity function and jet structure of Gamma Ray Bursts  }
\author[A. Pescalli et al.]
{A. Pescalli$^{1,2}$\thanks{E--mail:a.pescalli@campus.unimib.it}, 
G. Ghirlanda$^2$, O. S. Salafia$^{1,2}$ G. Ghisellini$^2$,
F. Nappo$^{3,2}$, R. Salvaterra$^4$ \\ \\
$^1$Dipartimento di Fisica G. Occhialini, Universita? di Milano Bicocca, Piazza della Scienza 3, I-20126 Milano, Italy\\
$^2$INAF -- Osservatorio Astronomico di Brera, via E. Bianchi 46, I-23807 Merate, Italy\\
$^3$Universita` degli Studi dellInsubria, via Valleggio 11, I-22100 Como, Italy\\
$^4$INAF -- IASF Milano, via E. Bassini 15, I-20133 Milano, Italy\\
}
\begin{document}

\date{}


\maketitle

\label{firstpage}
\begin{abstract}
The structure of Gamma Ray Burst (GRB) jets impacts on their prompt and afterglow 
emission properties.
The jet of GRBs could be {\it uniform}, with constant energy per unit solid angle 
within the jet aperture, or it could instead be {\it structured}, namely 
with energy and velocity that depend on the angular distance from the axis of the jet.
We try to get some insight about the still unknown structure of GRBs by studying their luminosity function. We show that low (10$^{46-48}$  erg s$^{-1}$) and high (i.e. with 
$L\ge 10^{50}$ erg s$^{-1}$) luminosity GRBs 
can be described by a unique luminosity function, which is also consistent with 
current lower limits in the intermediate luminosity range (10$^{48-50}$  erg s$^{-1}$). 
We derive analytical expressions for the luminosity function of GRBs in  
uniform and structured jet models and compare them with the data. 
Uniform jets can reproduce the entire luminosity function with reasonable
values of the free parameters. 
A structured jet can also fit adequately the current data, provided that
the energy within the jet is relatively strongly structured, i.e. 
$E\propto \theta^{-k}$ with $k\ge 4$. 
The classical $E\propto \theta^{-2}$ structured jet model is excluded by the current data. 
\end{abstract}

\section{Introduction}

Jets are a common feature of high energy astrophysical sources powered by accretion onto compact objects. 
GRB jets are thought to be the most extreme in terms of typical power 
( 10$^{50-54}$ erg s$^{-1}$) 
and of Lorentz factor ($\Gamma \sim 10^2 - 10^3$). 
Long GRBs are thought to follow the gravitational collapse of massive stars which form a 
rotating black hole with a bipolar jet along its rotational axis.  
The large isotropic equivalent energy $E_{\rm iso}$ of the prompt emission can  
exceed a solar mass rest energy, unless the radiation
is collimated (Tan, Matzner \& McKee 2001). 
The observed steepening of the afterglow flux light curve a few days after 
the burst is interpreted as the direct evidence of the presence of collimation. 

It has been typically assumed that GRBs have a \emph{Uniform Jet} (UJ): the energy and ejecta 
velocity are constant, within the jet aperture, 
and zero outside the jet opening angle (i.e. sharp--edged jet). 
During the afterglow phase, when the emission is produced by the deceleration of the relativistic jet 
by the interstellar medium,  a steepening of the observed afterglow flux is predicted when 
$\Gamma \sim 1/ \theta_{\rm j}$ (Rhoads 1997; Sari, Piran \& Halpern 1999). 
The measure of the time of this break ($t_{\rm break}$) has been used to infer the jet 
opening angle $\theta_{\rm j}$ which results distributed in the 1$^\circ$--10$^\circ$ range. 
Intriguingly, the true energetic of GRBs (i.e. accounting for their collimation) clusters 
around a typical value  \egamma$=$\eiso$(1-\cos\theta_{\rm j})\approx10^{51}$ erg  with a 
dispersion of less than a decade (Frail et al. 2001), correlating with the peak energy of the prompt emission spectrum 
$E_{\rm peak}$ (Ghirlanda et al. 2004; Nava et al. 2006).
Such typical value for the true energy of GRBs is also directly probed by late time radio 
observations (Frail, Waxman \& Kulkarni 2000; Frail et al. 2005; Shivvers \& Berger 2011).  

For small jet angles, the collimation corrected \egamma $\propto \theta_{\rm j}^2$\eiso.
The small dispersion of \egamma\ led to the idea that the jet is not uniform, but {\it structured},
where $\theta$ is the angular distance from the jet
axis, and coincides with the viewing angle $\theta_{\rm v}$.
This scenario assumes that what we believed to be the jet angle $\theta_{\rm j}$ 
is actually the viewing angle $\theta_{\rm v}$.
This can lead to a unification scheme in which all bursts are basically equal,
but appear different only because they are seen under different angles.
If the dependence of the burst energetics on $\theta$ is \eiso$\propto \theta^{-2}$,
one recovers the finding of Frail et al. (2001) of the clustering of \egamma.
%
These GRBs with a universal structured jet (SJ) were first proposed by 
Lipunov, Postnov \& Prokhorov (2001) and then studied by Rossi, Lazzati \& Rees (2002) and 
Zhang \& Meszaros (2002).
Some structure to the jet power and velocity could be imprinted,
within the collapsar model, by the interaction of the jet with the star 
(e.g. Zhang, Woosley \& Mac Fadyen 2003; Zhang, Woosley \& Heger 2004; 
Lazzati \& Begelman 2005; Morsony, Lazzati \& Begelman 2010; Levinson \& Eichler 2003; Lyutikov \& 
Blandford 2002) or instead the SJ could have an ``external" origin (Ghisellini et al. 2007).

Direct tests of the SJ model with available data, e.g. the search for a possible anti 
correlation between \eiso\ and \thv\ (Perna, Sari \& Frail 2003; 
Lloyd--Ronning, Dai \& Zhang 2004) or the statistical studies of the flux cumulative 
distribution of large GRB samples in the SJ model, did not provide definite evidence 
for this model (Nakar, Granot \& Guetta 2004; Cui, Liang \& Lu 2005). 
However, the presence of a structured jet is invoked to interpret some GRBs with a 
double jet break in their afterglow light curves (e.g. GRB 991216 -- Frail et al. 2000; GRB 030329 -- 
Berger et al. 2003, Sheth et al. 2003; GRB 021004 - Starling et al. 2005). 
In these cases, a ``discrete" SJ model composed by a narrow jet surrounded by 
a wider cone was proposed (Peng, Koenigl \& Granot 2005),  possibly due to the 
interaction of the jet with the star cocoon (Ramirez--Ruiz, Celotti \& Rees 2002; 
Vlahkis, Peng \& Koenigl 2003; Lazzati \& Begelman 2005). 

Several GRB properties are affected by the jet structure: the achromatic break \tjet\ 
in the afterglow light curve is smoother in the SJ model and it is related to
$\theta_{\rm v}$ rather than $\theta_{\rm j}$  
(Zhang \& Meszaros 2002); different degree of polarization 
of the afterglow emission (Rossi et al. 2004; Lazzati et al. 2004) and different 
luminosity functions (and GRB rates) are expected in the two scenarios. 
The jet break measurements are hampered by the necessity of a follow up of the afterglow
 emission until late times, by the smoothing induced by viewing angle effects 
 (e.g. van Eerten \& MacFadyen 2012) and by the contamination of the 
 afterglow emission at late times by the possible supernova and host galaxy 
 emission (e.g. Ghirlanda et al. 2007). 
Despite all these diagnostics, there is no concluding evidence yet of
the real jet structure.
On the other hand, 
the increasing number of bursts with measured redshift allows to estimate the
luminosity function of GRBs with increasing confidence.
We show in this paper how the jet structure affects the observed luminosity function.

The most recent studies of the GRB luminosity function are by 
Wanderman \& Piran 2010 with a relatively large GRB sample, and
by  Salvaterra et al. (2012) with a relatively small but complete GRB sample.
GRBs with measured redshift have typical isotropic equivalent luminosities in the 
range $10^{50-54}$ erg s$^{-1}$. 
At lower luminosities the detection rate drops to only a few events which are, however, 
representative of a large local density of GRBs (e.g. Soderberg et al. 2006; Pian et al. 2006). 
Indeed, it has been suggested that these latter events could belong to a different GRB population 
(e.g. Virgili et al. 2009; Daigne \& Mochkovitch 2007).

\subsection{Structure of this work}

In this paper we first consider the low luminosity GRBs to discuss if they can be accounted for  
by the extrapolation of the luminosity function (LF) describing high luminosity GRBs (\S2).
In this case we would have a luminosity function (LF) extending over 7 orders of
magnitude to be compared with different models.
Then we study four different possibilities:

\begin{enumerate}
\item
First we consider the simplest model: all jets are uniform, with the same $\theta_{\rm j}$ and are 
seen always on--axis, but we
assume that the jet had to punch the progenitor star before emerging, and in so doing
it must spend $\sim10^{51}$ erg. 
We derive the predicted LF in this case, and compare it to the data (\S3.1.1).

\item
Then we allow $\theta_{\rm j}$ to vary as a function of the observed luminosity, but we still 
assume that all bursts are seen on--axis.
Bursts of equal collimation corrected luminosity, but of different $\theta_{\rm j}$ will have
different isotropically equivalent luminosities $L_{\rm iso}$ (the smaller $\theta_{\rm j}$, the
larger $L_{\rm iso}$).
Also in this case we can construct analytically the LF, and we compare it with the data (\S3.1.2)

\item
We consider the possibility that (some) GRBs can be detected even if observed off--axis,
even if the jet is uniform. 
In fact any GRB will emit not only within $\theta_{\rm j}$, but also outside, even if the
corresponding radiation is not as relativistically boosted as the radiation inside the jet cone.
The observable off--axis luminosity depends on the bulk Lorentz factor
(the larger $\Gamma$, the dimmest the off--axis luminosity).
The predicted LF for a single value of $\Gamma$ and $\theta_{\rm j}$ can be derived analytically,
allowing for an easy numerical computation of the LF
with a distribution of $\Gamma$ and $\theta_{\rm j}$.
The latter is compared to the data (\S3.2).

\item
Finally, we consider structured jets, characterized by a power law dependence of the 
radiated energy with the angle from the jet axis ($E \propto \theta^{-s}$) to find out 
if the data are consistent with this scenario, and for which slope $s$ (\S4).

\end{enumerate}


We discuss our results in \S5. 
In the paper we assume a standard flat cosmology with $h=\Omega_{\Lambda}=0.7$. 

\section{Long GRB luminosity function}

The problem of deriving the luminosity function [LF -- $\Phi(L)$] of GRBs has been approached 
by different authors (Firmani et al. 2004; Guetta et al. 2005; Natarajan et al. 2005, Daigne et al. 2006, 
Salvaterra \& Chincarini 2007; Salvaterra et al. 2009; 2012)  by convolving $\Phi(L)$ with the GRB 
formation rate $R_{GRB}(z)$ (proportional to the cosmic star formation rate). 
The free parameters of the LF  can be constrained by fitting the resulting model to the flux 
distribution of large GRB samples (e.g. the \cgro/Batse population). 
Note that, in the GRB literature, it is customary to consider as cosmic evolution any
(i.e. density, or luminosity, or both) evolution {\it in addition} to the one related to 
the evolution of the star formation rate.
The main difficulty of these studies is accounting, in the model, for the selection effects which 
affect the true GRB population detected by any instrument. 
Recently, Salvaterra et al. (2012 -- S12 hereafter) 
constructed a flux limited sample of bright GRBs detected by \sw/BAT  which resulted 95\% complete in redshift. 
While the possible evolution of the  luminosity function or of the GRB formation rate with redshift 
is still a matter of debate, there seems to be a general consensus on the shape of the LF: 
\begin{equation}
\Phi(L) \propto
\begin{cases}
\left( \frac{L}{L_{\rm c}} \right)^{-a},  & \text{ $L \leq L_{\rm c}$} \\
\left( \frac{L}{L_{\rm c}} \right)^{-b} ,  & \text{ $L > L_{\rm c}$}
\end{cases}
\label{eq1}
\end{equation}
where $L_c$ represents the break luminosity. S12 finds $a=1.56^{+0.11}_{-0.42}$, 
$b=2.31^{+0.35}_{-0.31}$, $L_{c}=2.5^{+6.8}_{-2.1}\times 10^{52}$ erg s$^{-1}$ 
(68\% confidence intervals) for the case of no evolution of $\Phi(L)$.


\begin{figure*}
\hskip -1.2truecm
\psfig{file=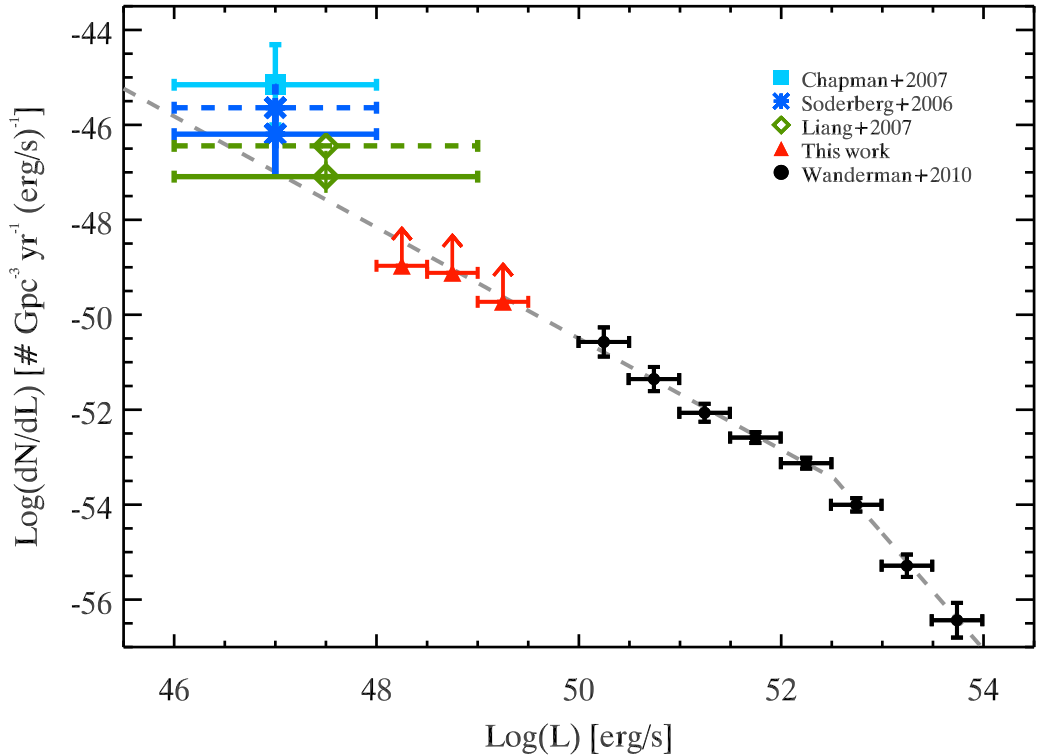,width=16cm}
\vskip -0.3 cm
\caption{
Long GRB luminosity function representing the number of GRBs per unit volume, 
time and luminosity bin. 
Black symbols represent the discrete luminosity function of WP10. 
The grey dashed line is the empirical fit of WP10 with a broken power law  
with $a=1.2$, $b =2.4$ and $L_{\rm c}=3.2\times10^{52}$ erg s$^{-1}$.
The rate of low luminosity GRBs is shown with different symbols according to the different 
sources in the literature: 
asterisk (Soderberg et al. 2006), diamond (Liang et al. 2007), filled square (Chapman et al. 2007). 
The Soderberg et al. (2006) and Liang et al. (2007) rates were calculated assuming a time bin corresponding 
to the \sw\ lifetime in 2006 (points with dashed horizontal bars).
Since then, no other burst in the same luminosity bin has been discovered,  
so we added the rates corrected for the current \sw\ lifetime (points shown with solid horizontal bars). 
The lower limits on the rate of intermediate luminosity GRBs (IL) added in this work are 
shown with the filled (red) triangles.
}
\label{fg1}
\end{figure*}

\subsection{High Luminosity (HL) GRBs}

Wanderman \& Piran (2010) (WP10 hereafter) adopted a direct inversion method on the distribution of 
GRBs in the $L-z$ space obtaining simultaneously $\Phi(L)$ and $R_{GRB}$ 
independently\footnote{This method relies on the assumption of no evolution of the GRB luminosity 
function and rate with redshifts.
See WP10 for the validity of this assumption.}. 
They selected a sample of 
long\footnote{See Wanderman \& Piran (2014) for the same method applied to short GRBs.} 
GRBs with spectroscopically measured redshift and isotropic equivalent luminosities $L_{\rm iso}\ge 10^{50}$ 
erg s$^{-1}$ detected by BAT on board \sw. 
The derived LF is represented by a discrete series of data points (Fig. 3 of WP10) in 
eight equal logarithmic bins of luminosity and can be represented by a 
broken power law 
with $a=1.2^{+0.2}_{-0.1}$ and $b=2.4^{+0.3}_{-0.6}$, 
with the break at $L_{\rm c}=10^{52.5\pm 0.2}$ erg s$^{-1}$ 
(note that WP10 use $dN/d\log L$, whereas we prefer to adopt $dN/dL=dN/(L\,\, d\log L$) 
so that the WP10 slopes are here increased by 1).

These parameter values are consistent with those derived with the ``classical" approach described above. 
We  normalized the luminosity function 
at the local GRB rate $\rho_{0} \simeq 1.3$ Gpc$^{-3}$ yr$^{-1}$ (WP10). 
Fig. \ref{fg1} shows the data points of WP10 (black symbols) which cover the luminosity 
range between 10$^{50}$ and 10$^{54}$ erg s$^{-1}$ and will be referred to as HL bursts hereafter. 
The best fit obtained by WP10 is shown as a grey long dashed line.

\subsection{Low Luminosity  (LL) GRBs}

At the low end of the luminosity distribution of GRBs, i.e. $L_{\rm iso}\sim10^{46-48}$ erg s$^{-1}$,  
there are two events (GRB 980425 and GRB 060218) which have been detected in the local Universe and have been intensively 
studied as direct evidences of the massive star progenitor of long GRBs. 
Their luminosity 
is three orders of magnitude smaller than HL bursts, and their rate is larger (e.g. Soderberg et al. 2006). 
GRB 980425 ($z=0.008$, associated to SN1998bw -- Galama et al. 1998) was detected by \cgro/Batse 
and had $L_{\rm iso}\sim7\times10^{46}$ erg s$^{-1}$ (as computed from its prompt emission 
spectrum -- Jimenez, Band \& Piran 2001).
Similarly, GRB 060218 ($z=0.0331$, associated to SN2006aj -- Sollerman et al. 2006), 
detected by \sw/BAT, had $L_{\rm iso}\sim 1.3\times10^{47}$ erg s$^{-1}$ (Campana et al. 2006) . 

The rate of these LL events can be computed as 
\begin{equation}
\rho_{LL}\simeq 4\pi { N_{\rm LL}\over V_{\rm max} T \Omega}
\end{equation}
where $V_{\rm max}$ is the maximum volume
 within which they could be detected by an instrument with an assigned sensitivity, with a field 
 of view $\Omega$ and operating for a time $T$. 
 Based on the two GRBs 980425 and 060218, Soderberg et al. (2006; see also Pian et al. 2006) 
 derived the rate of LL events  by conservatively averaging 
 over $V_{\rm max}$ and $\Omega$ for different detectors ({\it Beppo}SAX/WFC, {\it Hete--II}/WXM and \sw/BAT). 
 They obtained a rate $\rho_{\rm LL}\sim 230^{+490}_{-190}$ Gpc$^{-3}$ yr$^{-1}$. 
 In the luminosity range 
 10$^{46}$--10$^{48}$ erg s$^{-1}$ occupied by these two GRBs and centered at $\langle L\rangle=10^{47}$ 
 erg s$^{-1}$ we convert this rate dividing it for the interval width obtaining $\tilde{\rho}_{\rm LL} \sim 2.3\times10^{-46}$ Gpc$^{-3}$ 
 yr$^{-1}$ erg$^{-1}$ s. 
 This is represented by the (blue) asterisk in Fig. \ref{fg1}. 
 Since $\rho_{\rm LL}$ 
 has been computed in 2006, we have also corrected it (blue asterisk with solid horizontal bar in Fig. \ref{fg1}) for the larger 
 time interval elapsed since the detection 
 of these two LL events. 
 These results  are  consistent with numerical studies: Virgili, Liang \& Zhang (2009) estimate 
 $\rho_{LL}$=200  Gpc$^{-3}$ yr$^{-1}$ for events with $\langle L\rangle=10^{47}$ erg s$^{-1}$ 
 based on the Batse GRB population. 
A slightly larger rate $\rho_{\rm LL}\sim$700$\pm$360 Gpc$^{-3}$ yr$^{-1}$ 
(shown by the cyan filled square symbol in Fig. \ref{fg1}) has been obtained by 
Chapman et al. (2007) from the cross--correlation of 
a subsample of low--fluence smooth single--peaked Batse bursts with nearby galaxies. 
Liang et al. (2007) also derived $\rho_{\rm LL}= 325^{+352}_{-177}$ Gpc$^{-3}$ yr$^{-1}$ 
(shown by the green diamond symbol in Fig. \ref{fg1} and also corrected for the lifetime of \sw).

\begin{table*}
\begin{center}
\begin{tabular}{llllllllllllll}
\hline\hline
GRB & $z$ &Ref$^{a}$ &$\alpha$ &$\beta$ &$E_{\rm peak}$ &$P$ &$\Delta E$ &Ref$^b$ & $P_{\rm bol}$ &$L_{\rm iso}$ &Instr. &$P_{\rm lim,bol}$ \\  
    &     &          &         &        &keV            &ph/cm$^2$/s    &keV        &  & ph/cm$^2$/s &erg/s  &       &ph/cm$^2$/s \\
\hline
020903   &0.25   &1 &--1.0  &...   &3.37        &2.8  &[2--400]   &7 & 6.52 &$4.86\times10^{48}$ &\hete          &3.0 \\
031203   &0.105  &2 &--1.63 &...   &144         &2.2  &[15--150]  &8,9,10 & 17.9 &$10^{49}$           &{\it Integral} &3.0 \\
051109B  &0.08   &3 &--1.90 &...   &50$^{\dag}$ &0.5  &[15--150]  &11,12,13 & 9.43 &$1.64\times10^{48}$ &\sw            &1.3 \\
060505   &0.089  &4 &--1.8  &...   &239$^{\dag}$ &1.9  &[15--150]  &11 & 8.0  &$7.14\times10^{48}$ &\sw            &0.8\\
120422A  &0.283  &5 &--1.94 &...   &53          &0.6  &[15--150]  &14 & 11.35  &$2.7\times10^{49}$  &\sw            &2.5\\
130702A  &0.145  &6 &--1.0  &--2.5 &20          &7.03 &[10--1000] &15 & 7.03  &$2.87\times10^{49}$ &\sw            &2.5  \\
\hline\hline
\end{tabular}
\end{center}
\caption{
Intermediate Luminosity (IL) GRBs. $^{a}$ References for the redshift: 
(1) GCN \#1554 Soderberg, Price, Fox et al. (2002); (2)  GCN \#2482 Prochaska, Bloom, Chen, Hurley, Dressler $\&$ Osip (2003); (3)  GCN \# 5387 Perley, Foley, Bloom $\&$ Butler (2005) ; (4)  GCN \#5161 Thoene, Fynbo, Sollerman et al. (2006); (5)  GCN \#13251 Tanvir, Levan, Cucchiara et al. 2012 ; (6)  GCN \#14983 Leloudas, Fynbo, Schulze et al. (2013); 
$^{b}$ References for the spectral parameters: 
(7) Sakamoto et al. (2004) ; (8) Bosnjak et al. (2013); (9) Sazonov et al. (2004); (10) Ulanov et al. (2005); 
(11) Troja et al. (2006); (12) Sakamoto et al. (2009); (13) Butler et al. (2007); (14) Zhang et al. (2012); (15) Kienlin et al. (2014). $\dag$ $E_{\rm peak}$ computed through the $\alpha-E_{\rm peak}$ correlation of 
Sakamato et al. (2009) for \sw\ GRBs (for GRB 051109B consistent also with the estimate of Butler et al. 2007).
}
\label{tab1}
\end{table*}%

\subsection{Intermediate Luminosity (IL) GRBs}

In the intermediate luminosity range between HL and LL (see Fig. \ref{fg1}) we can add some constraints. 
We have searched all GRBs with measured 
redshift\footnote{http://www.mpe.mpg.de/$\sim$jcg/grbgen.html} 
$z$ and with $L_{\rm iso}\in [10^{48},10^{50}]$ erg s$^{-1}$. 
In addition to $z$, we also required that $L_{\rm iso}$ is well determined: this is possible 
when the spectrum at the peak of the light curve has been fitted over a wide enough energy range as 
to constrain its peak energy. 
Indeed, in these cases it is possible to compute the bolometric isotropic luminosity. 
We adopted the same method described for LL 
(from Soderberg et al. 2006) to compute $\rho_{\rm IL}$ in three luminosity bins,
using the following bursts: 
(i) GRB 051109B with $L_{\rm iso}\in[10^{48},3\times10^{48}$ erg s$^{-1}$], 
(ii) GRB  020903, 031203, 060505 with $L_{\rm iso}\in[3\times10^{48},10^{49}$ erg s$^{-1}$]
and (iii) GRB 120422A, 130702 with $L_{\rm iso}\in[10^{49},3\times10^{49}$ erg s$^{-1}$]. 

We have collected the prompt emission spectral parameters and flux of these bursts 
(reported in Tab. \ref{tab1} -- Col. 4--8) through which we have computed their bolometric flux  
(Col. 11 in Tab. \ref{tab1}). 
Three different instruments were involved in triggering these 
events (Col. 10 in Tab. \ref{tab1}) and we considered the following instrumental parameters: 
$\Omega=1.33$ sr and $T=8$ yr for \sw, $\Omega=0.1$ sr and $T=10$ yr for {\it Integral} 
and $\Omega=0.802$ sr and $T=4$ yr for \hete. 
The last column in Tab. \ref{tab1} reports the limiting flux of the corresponding detectors as computed by Band (2002; 2006) which 
depends on the burst peak spectral energy $E_{\rm peak}$ (in the observer frame). 
$P_{\rm lim}$ as computed by Band (2002; 2006) in the [1-1000] keV observed energy band is used to compute the maximum distance (and therefore $V_{\rm max}$) out to which these events could have been detected. 
With the same method adopted for LL bursts we could derive a rate $\rho_{IL}$ for intermediate luminosity events. 
These rates should be considered as lower limits: we only selected GRBs with measured redshifts and 
well constrained spectral parameters. These are most likely only a fraction of the bursts, with similar luminosities, 
which effectively triggered the corresponding detector. These rates are shown by the (red) triangles in Fig. \ref{fg1}.

Finally, Fig. \ref{fg1} shows the luminosity function of HL bursts (data points from WP10 -- black symbols), 
IL bursts (lower limits -- red triangles) and LL bursts (colored symbols - references in the caption). 
The grey dashed line represents the LF fitted by WP10 to their data points (only HL bursts) and it 
can be noted that its extrapolation to low luminosities is consistent with both the lower limits 
of IL bursts and the rate of LL events. 
This is a direct indication that LL and  HL have a common progenitor, i.e. they form a \emph{unique population}. 
Apparently there is no need to invoke a different origin for the LL events as they are consistent with 
the extension to low luminosities of the LF of HL bursts. 
Based on this evidence, in the following we compare the empirical LF with model predictions for different 
jet configurations. 

\begin{figure*}
\hskip -1.2truecm
\psfig{file=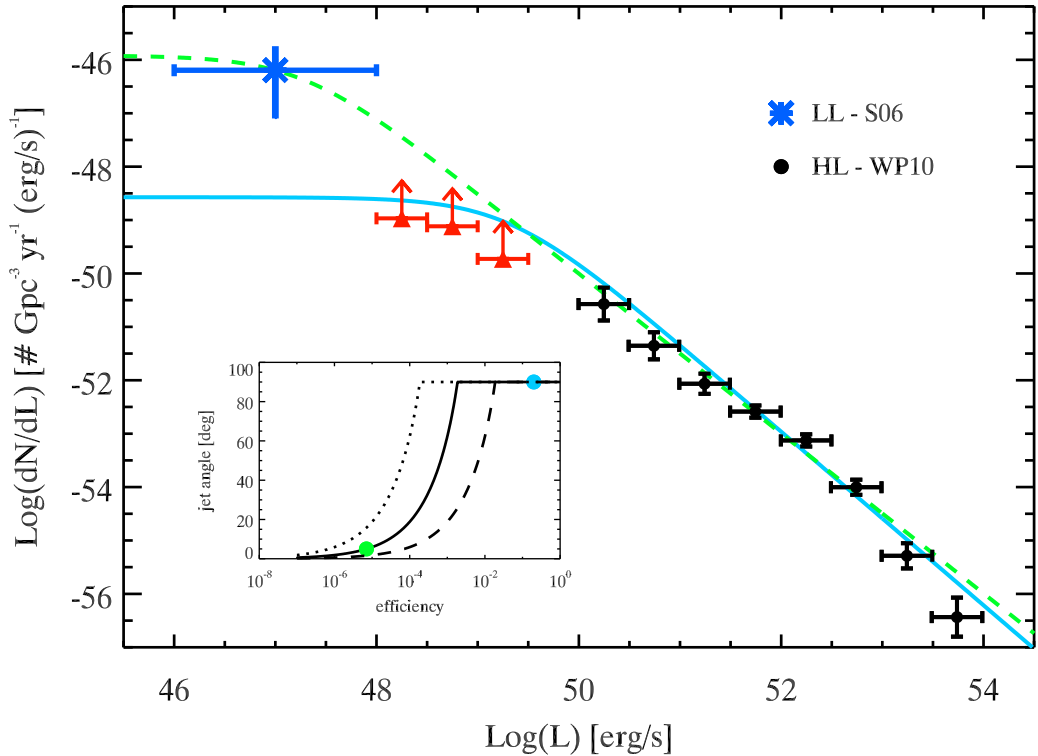,width=16cm}
\vskip -0.3 cm
\caption{
Luminosity function fitted with the UJ model (Eq. \ref{uj}) with fixed opening angle. 
For the models shown in the main plot (solid cyan and dashed green curves) $L_{\star}=10^{50}$ erg s$^{-1}$ is fixed. 
The solid cyan line shows the case with $\eta=0.2$ (which gives \th=90$^{\circ}$) and the dashed green line 
is the case with \th=5$^{\circ}$ (which gives $\eta\sim10^{-5}$). 
{\it Inset}: curves showing the degeneracy of the model in the angle and efficiency parameters. 
Dotted, solid and dashed curves correspond to different choices of 
$L_{\star}=10^{49}, 10^{50}, 10^{51}$ erg s$^{-1}$, respectively. 
The green and cyan dots show the choice of $\eta$ and \th\ corresponding to the model curves 
shown in the main panel, respectively.  
}
\label{fg2}
\end{figure*}

\section{Uniform jet}

In this section, we derive the analytic expression of the luminosity function under 
the assumptions of UJ, e.g. we assume that 
the energy per unit solid angle $\epsilon$ and the initial bulk Lorentz factor \G\ 
are constant within the jet, and zero outside. 

\subsection{Jet break--out energy }

To be observed,  the jet must 
emerge from the progenitor star.
This implies that a minimum central engine duration is required for a jet to successfully 
escape the progenitor star. 
Bromberg et al. (2012) derived 
interesting consequences from this simple consideration, concerning
the GRB duration distribution.
If the central engine lasts much longer than the time necessary for the jet to drill through the 
star (break out time), the resulting GRB  will be a long one. 
If the central engine duration is comparable to the break out time,
but slightly longer,
the GRB will be  short. 
As a consequence Bromberg et al. (2012, 2013) estimated the  fraction of short GRBs produced 
by collapsars and revised the separation of long/short events. 


The same argument can be applied to the energetics. In a recent work Kumar $\&$ Smoot (2014) find that the minimum energy required to excavate the star envelope must be at least the energy contained in the cocoon. Based on M\'esz\'aros $\&$ Rees (2001) we assume that the energy associated to the cocoon is $E_{\star} \simeq 10^{51}$ erg. 

Therefore, if the inner engine provides enough energy $E_{\rm kin}>E_{\star}$, the jet can escape 
the star and the GRB can be observed. 
The ``residual" energy of the burst will be $E_{\rm kin}-E_{\star}$. 
If $E_{\rm kin}\gg E_{\star}$, a normal HL burst is observed, while LL events are those with 
$E_{\rm kin}$ only slightly larger than $E_\star$.
We show below that, likewise to the duration distribution (Bromberg et al. 2012), 
also the energy distribution should be flat below the characteristic $E_{\star}$. 

Assume that the central engines of GRBs provide a total kinetic energy distributed 
as $P(E_{\rm kin}) = dN(E_{\rm kin})/dE_{\rm kin} \propto E_{\rm kin}^{-k}$.
Of the total kinetic energy $E_{\rm kin}$ produced by the central engine, only  
$E_{\rm kin}-E_{\star}$ is available, left after the jet has escaped the progenitor star.
Moreover, only {\it a fraction} $\eta$ of the residual kinetic energy $E_{\rm kin}-E_{\star}$ can be 
converted into radiation ($\eta$ is typically a few per cent in the standard internal shock model -- Rees \& Meszaros 1994). 

If $E_{\gamma}=\eta(E_{\rm kin}-E_{\star})$ is the energy converted into radiation, 
we can define its isotropic equivalent as: 
\begin{equation}
E_{\rm iso} = \frac{\eta(E_{\rm kin} - E_{*})}{1-\cos\theta_{\rm j}}
\label{eiso}
\end{equation} 
where $\theta_{\rm j}$ is the jet opening  angle. 
The relation between $P(E_{\rm kin})$ and $P(E_{\rm iso})$ is:
\begin{equation}
P(E_{\rm iso}) =  P(E_{\rm kin}){ dE_{\rm kin} \over dE_{\rm iso}}
\label{peiso}
\end{equation} 
%
To pass from the energy function to the luminosity function we take advantage of the fact
that the distribution of the rest frame duration of GRBs peaks at $25$ s.
This is true also for the bursts used by WP10. 
Assuming that the light curve of the prompt emission can be approximated by a triangular shape, 
we have $E_{\rm iso}\approx (t \times L_{\rm iso})/2$.

Then the luminosity function in the UJ model is:
\begin{eqnarray}
P(L_{\rm iso}) \, &=& P(L_{\rm kin}) { dL_{\rm kin} \over dL_{\rm iso} } = L_{\rm kin}^{-k} 
{ 1-\cos\theta  \over \eta}  \nonumber \\
&\propto& \frac{(1-\cos\theta_{\rm j})}{\eta}
\left[ {L_{\rm iso} (1-\cos \theta_{\rm j}) \over \eta} + L_{\star}\right]^{-k}
\label{ujall}
\end{eqnarray} 

where $L_\star$ is the collimation corrected (i.e. ``true") {\it kinetic} luminosity
necessary to punch the star, while $L_{\rm iso}$ is the observed, isotropically equivalent
{\it radiative} luminosity.
Eq. \ref{ujall} is the LF of {\it all} bursts, including the ones not pointing at us.
If all bursts have the same $\theta_{\rm j}$, we will see only a fraction $(1-\cos\theta_{\rm j})$
of them, independent of luminosity.
Therefore the observed LF is: 
\begin{equation}
P(L_{\rm iso}) \propto{(1-\cos\theta_{\rm j})^2 \over \eta}
\left[ {L_{\rm iso} (1-\cos \theta_{\rm j}) \over \eta} + L_{\star}\right]^{-k}
\label{uj}
\end{equation}
Only those engines providing $L>L_{\star}$ will produce a successful GRB and build up the 
$P(L_{\rm iso})$ distribution: if $L\gg L_{\star}$ then  $P(L_{\rm iso})\propto L_{\rm iso}^{-k}$, 
whereas for $L \lsim L_{\star}$ the luminosity function is flat:
\begin{equation}
P(L_{\rm iso}) \propto
\begin{cases}
{\rm const},      & \text{ $L_{\rm iso} \ll  \eta L_\star /(1-\cos\theta_j)$} \\
L_{\rm iso}^{-k}, & \text{ $L_{\rm iso} \gg   \eta L_\star /(1-\cos\theta_j)$} 
\end{cases}
\label{uj2}
\end{equation}
The transition between these two regimes corresponds to a characteristic luminosity, i.e. 
$L_{\rm iso}\sim \eta L_\star/(1-\cos\theta_{\rm j})$.


\subsection{UJ observed in--jet [\thv$<$\th]}

Independently from the jet structure (UJ or SJ), one important parameter affecting the 
observed properties of GRBs is the viewing angle \thv, i.e. the angle between the observer 
line of sight and the jet axis. 
In the UJ model, we first assume that the emission can be seen only if the observer line of 
sight intercepts the jet aperture angle, i.e. \thv$\le$\th. 
This is a good approximation if the bulk Lorentz factor of the prompt phase $\Gamma_0$ is 
relatively large, i.e. $1/\Gamma_0\ll$\th.

\subsubsection{Unique jet opening angle \th}

Assume that all GRBs have the same \th. 
$P(L_{\rm iso})$ in Eq. \ref{uj} represents the luminosity function of the {\it observed} GRBs. 
The {\it total} LF is given in Eq. \ref{ujall}.
The factor $(1-\cos\theta_{\rm j})$ is in this case constant, and can be
absorbed in the normalization of Eq. \ref{uj}. 
Fig. \ref{fg2} shows the fit of the luminosity function with the model of Eq. \ref{uj}.  
We fit the HL rates (black symbols) and the LL rate (blue asterisk) as derived by Soderberg et al. 
(2006 -- corrected in this work for the elapsed time, see \S2). 
The lower limits of IL bursts (red triangles in Fig. \ref{fg2}) are used only for 
a consistency check of the fitted model. 

Fixing $L_{\star}=10^{50}$ erg s$^{-1}$, the free parameters are 
the normalization, the slope $k$
and the characteristic luminosity  
$L_{\star}\eta/(1-\cos\theta_{\rm j})$. 
Since the model depends on the ratio between $\eta$ and $(1-\cos\theta_{\rm j})$,
there is degeneracy between these two quantities.
The fit can constrain this ratio rather than the two factors independently. 
We consider two cases: 
\begin{itemize}

\item 
$\eta$=0.2 (as typically found from the modeling of the GRB afterglows -- e.g. Panaitescu 
\& Kumar 2002): the fit is shown by the solid cyan line in Fig. \ref{fg2}. 
HL bursts can be reproduced with a unique LF which has a slope $k=1.62\pm0.08$ 
(1$\sigma$ confidence) and is also marginally consistent with the IL lower limits. 
The fit has $\chi^2=33.5$ for 6 degrees of freedom (dof) corresponding to a goodness 
of fit probability of 8$\times 10^{-6}$. 
However, the LL bursts cannot be reproduced by this model. 
Indeed, the characteristic luminosity $L_{\star}\eta/(1-\cos\theta_{\rm j})$, with 
$\eta=0.2$ and $L_{\star}=10^{50}$ erg s$^{-1}$, corresponds to 
$\sim2\times 10^{49}/(1-\cos\theta_{\rm j})$ erg s$^{-1}$. 
This expression has a minimum for \th=90$^\circ$. 
In this case all GRBs would be isotropic and still LL events should be a different population; 

\item 
\th=5$^\circ$ (i.e. corresponding to the typical  opening angle of GRBs -- 
Frail et al. 2001; Ghirlanda et al. 2007): the fit is shown by the 
dashed green line in Fig. \ref{fg2} and can reproduce all the bursts (with a $\chi^2=16$ for 6 dof, 
i.e. probability 0.01). 
The resulting $k=1.49\pm0.08$ is consistent with the previous one,  but the  efficiency 
$\eta\sim 10^{-5}$ is unreasonably low.  
\end{itemize} 

We have verified that the above results are independent from the choice of the particular 
value for $L_{\star}$: if we fix this parameter to any value  $[10^{49},10^{51}]$ erg s$^{-1}$, 
we still find an unreasonably low efficiency $\eta<10^{-5}$ for ``reasonable" \th; or too large \th\ for ``reasonable" efficiencies.
This is shown in the inset of Fig. \ref{fg2} where all the curves saturate at 
90$^\circ$ (for the three different choices of $L_{\star}$) for $\eta>0.01$.

We conclude that the UJ model (assuming a unique angle for all bursts) does not reproduce 
the entire LF (from LL to HL bursts).

\begin{figure*}
\hskip -1.2truecm
\psfig{file=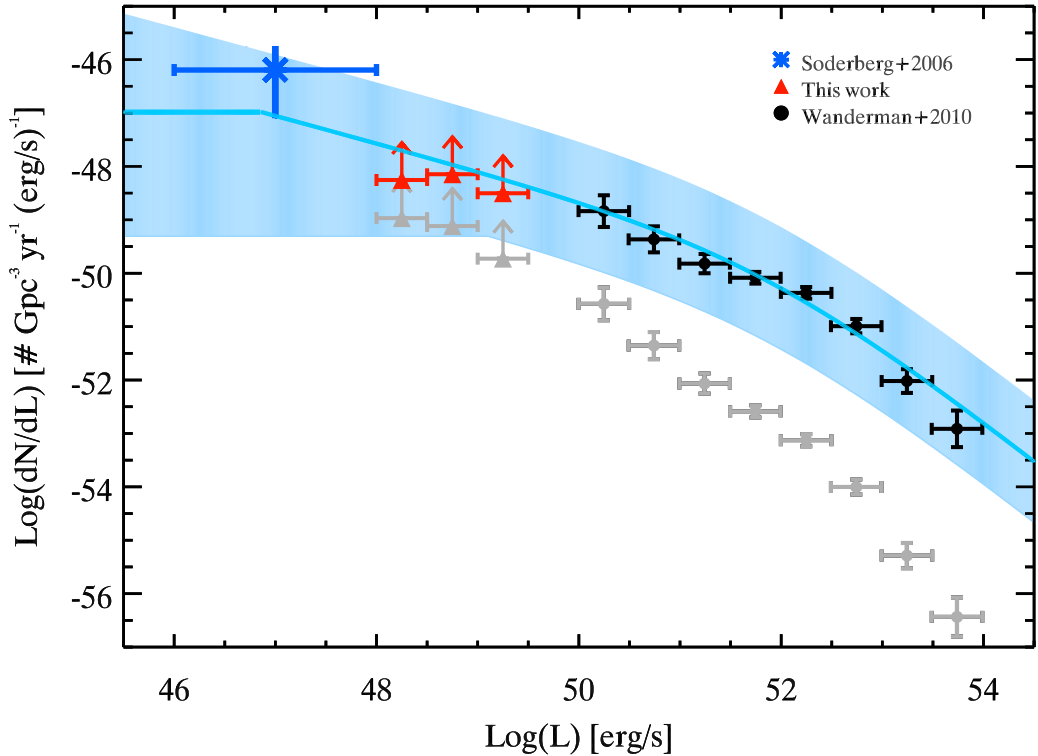,width=16cm}
\vskip -0.3 cm
\caption{
Uniform Jet with jet angle dependent from the luminosity. 
The LF of the entire GRB population (i.e. corrected for the collimation factor, 
which depends on the luminosity in this case) is shown by the black symbols. 
Original points (grey symbols) are also shown. 
The model is shown by the solid cyan line and the 3$\sigma$ confidence interval of the model, 
obtained accounting for the scatter of the \th--$L_{\rm iso}$ correlation (see text) 
is shown by the azure shaded region.
}
\label{fg3}
\end{figure*}

\subsubsection{Jet angle depending on luminosity  [$\theta_{\rm j}(L)$]}

The assumption made in the previous section, that all GRBs have the same 
jet angle, may be relaxed.
Indeed there is the possibility
that the LL bursts have wide opening angles to account for their low luminosity. 
If the luminosity range of the LF is due to the range of opening angles, one should 
expect an anti correlation between $L_{\rm iso}$ and \th. Indeed, it has been shown that 
such a relation exists for GRBs with measured \th\ (e.g. Lloyd--Ronning et al. 2004; 
Firmani et al. 2005; Ghirlanda et al. 2005). 

The peak energy $E_{\rm peak}$ of GRBs is correlated with the isotropic energy 
$E_{\rm iso}$ as \ep$=k_{A} E_{\rm iso}^{A}$ (Amati et al. 2002) with $A\sim0.5$ 
(a similar correlation exists with the isotropic luminosity:
\ep$=k_{Y} L_{\rm iso}^{Y}$ -- Yonetoku et al. 2004). 
The slopes of these two correlations are similar $A\sim Y \sim 0.5$.

If $E_{\rm iso}$ is corrected  for collimation, namely
$E_\gamma= E_{\rm iso} (1-\cos\theta_j)$, 
a tighter and steeper 
correlation \ep$=k_{G} E_{\gamma}^G$ (Ghirlanda et al. 2004) is found. 
The slope depends on the density profile of the circumburst medium,
being $G=0.7$ for a homogeneous density 
and $G=1$ for a wind profile (Nava et al. 2006). 
Since the \ama\ and \ghi\ correlations differ for the collimation 
factor (i.e. $1-\cos\theta_{\rm j}$), their different slopes  suggest that there is 
a relation between the average jet opening angle and the energy of GRBs: more 
energetic bursts should have a smaller \th. 

We can derive this relation by combining the correlations: 
\begin{equation}
1-\cos\theta_{\rm j}=
\left(\frac{k_A}{k_G}\right)^{\frac{1}{G}}E_{\rm iso}^{\frac{A-G}{G}} = C L_{\rm iso}^{-\xi}
\label{amaghi}
\end{equation} 
where $\xi=(G-A)/G$.  
The typical duration, assumed $t=25$ s (rest frame) to 
convert energetics to luminosities, has been incorporated in the normalization constant $C$. 
This relation 
also establishes that there is a 
characteristic minimum luminosity  $L_{\rm iso,min}\sim 7\times10^{46}$ erg s$^{-1}$ 
corresponding to $\theta_{\rm j}=90^\circ$. 

Since $\theta_{\rm j}$ is not constant, it is convenient to find the 
LF of all bursts, including the ones not pointing at us, i.e. using Eq. \ref{ujall}.
The luminosity function then becomes: 
\begin{equation}
P(L_{\rm iso}) 
\propto {C(1-\xi)L_{\rm iso}^{-\xi}\over \eta}\left(\frac{CL_{\rm iso}^{1-\xi}}{\eta} 
+ L_{\star}\right)^{-k}
\label{amaghiLF}
\end{equation} 
where the factor $(1-\xi)$ comes from the derivative $dL_{\rm kin}/dL_{\rm iso}$
in which now also the collimation factor depends on $L_{\rm iso}$.
At low $L_{\rm iso}$, $N(L_{\rm iso}) \propto L^{-\xi}$, 
while at large  $L_{\rm iso}$, $N(L_{\rm iso})\propto L^{-\xi-k(1-\xi)]}$.
The break is at $L_{\rm iso}\sim (\eta L_\star/C)^{1/(1-\xi)}$.

In Fig. \ref{fg3} the corrected points (black dots and red triangles) are shown 
together with the observed ones (grey dots): 
the higher the luminosity, the larger is the applied correction, according to Eq. \ref{amaghi}. 
Below a limiting $L_{\rm iso}$ the opening angle becomes larger than $90^\circ$ and
we do not apply the correction any longer.
Since smaller angles correspond to larger $L_{\rm iso}$, the LF of all GRBs 
is flatter than the observed one.

Fixing $L_{\star}=10^{50}$ erg s$^{-1}$, the fit favors an efficiency 
$\eta=1$ (pegged to its maximum value) which is  unrealistic (e.g. there would be 
no energy left for the afterglow emission). 
Similarly to the previous case, changing $L_{\star}$ by a factor 
10 still gives $\eta=1$. 
This is due to the degeneracy between $\eta$ and $L_{\star}$. 
Only for $L_{\star}\ge10^{52}$ erg s$^{-1}$ we find reasonable values of $\eta$. 
Therefore, we decided to show in Fig. \ref{fg3} the fit with both $\eta$ and $L_{\star}$ fixed to their 
typical values (0.2 and $10^{50}$ erg s$^{-1}$ respectively) (cyan solid line). 
The fit is consistent with the lower limits of the IL bursts (triangles in Fig. \ref{fg3}) 
and marginally consistent with the LL bursts. 
Setting $A=0.5$, $G=1$, we have $\xi=0.5$, which is the slope at low luminosities.
The slope of the distribution of energies provided by the inner engine found by the fit is $k = 1.5\pm0.15$, corresponding to an high luminosities slope $\xi+k(1-\xi) \simeq 1.25$ for $\Phi(L)$.
We stress that ``finding the best fit" is here only a formal procedure:
to the errors associated with the data points we should in fact add the uncertainties
about the \ama\ and \ghi\ correlations, that have a scatter of 0.22 dex and 0.12 dex, respectively. 
Fig. \ref{fg3} shows (shaded region) the 3$\sigma$ boundaries of the LF obtained accounting for the 
scatter of these correlations in deriving the correlation of Eq. \ref{amaghi}. 
The LL point is at the border of this confidence interval.

\begin{figure*}
\hskip -1.2truecm
\psfig{file=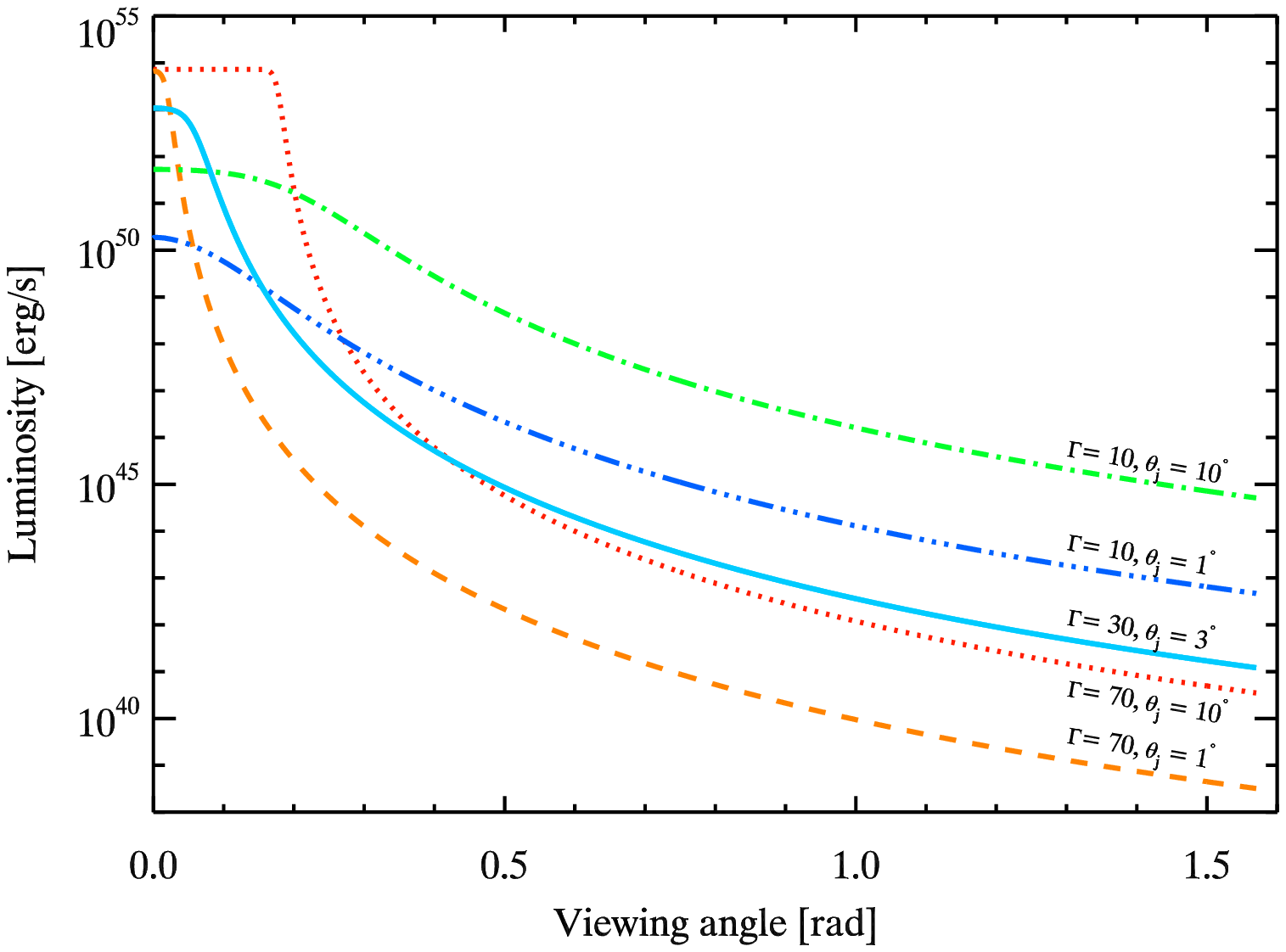,width=9cm}
\psfig{file=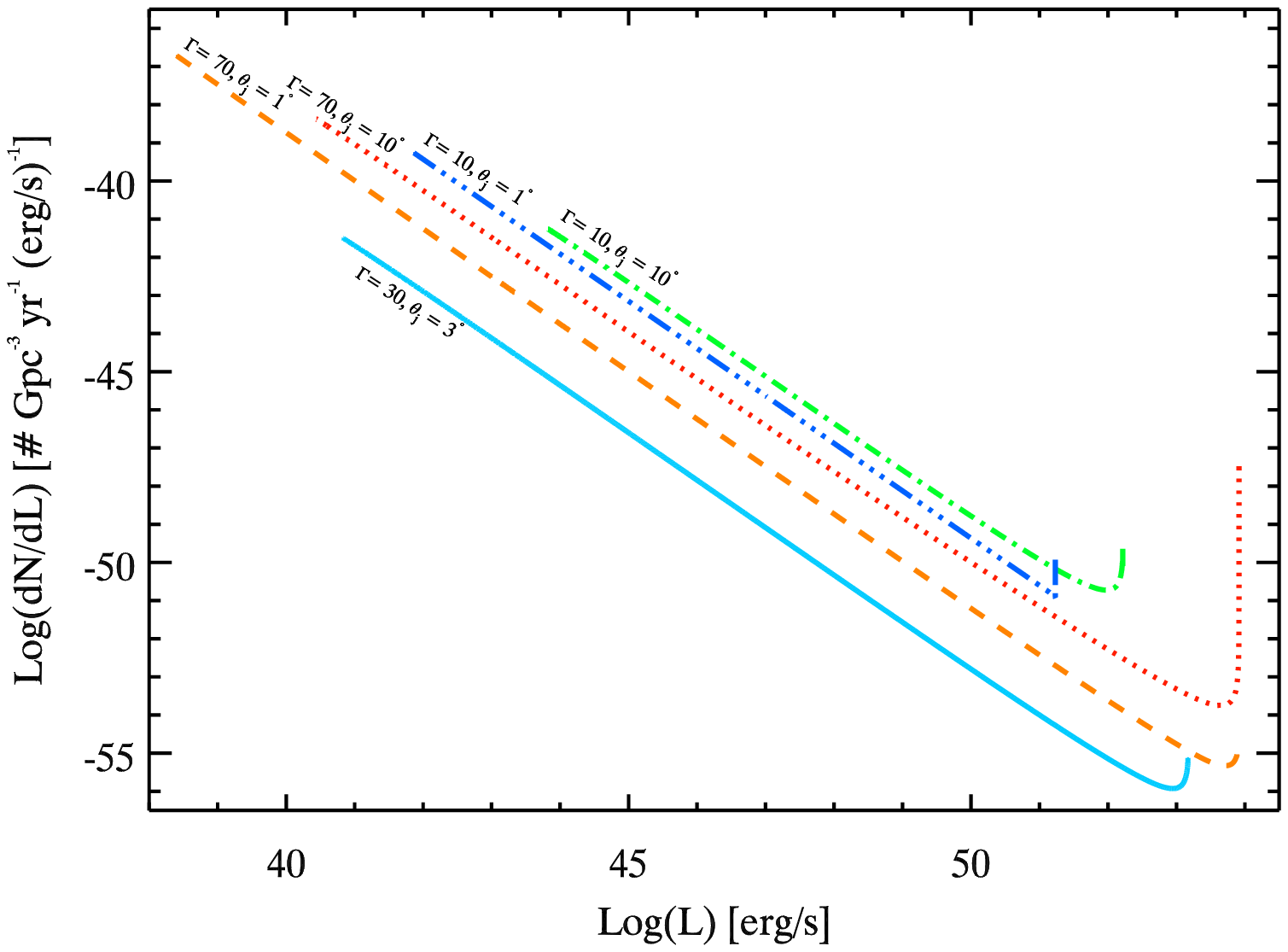,width=9cm}
\vskip -0.3 cm
\caption{
{\it Left panel}: luminosity as a function of the viewing angle  for 
different combinations of the bulk Lorentz factor $\Gamma$ and jet opening angle \th. 
The comoving frame luminosity $L^\prime$ is the same for all curves. 
The different lines (colors) show different assumed combinations of \th\ and $\Gamma$. 
For all the curves we have assumed $L^\prime=10^{49}$ erg s$^{-1}$. 
{\it Right panel}: Luminosity functions corresponding to the different 
assumptions of \th\ and $\Gamma$ of the right panel. 
}
\label{fg4}
\end{figure*}

\subsection{UJ observed in-- and out--jet [\thv$\lessgtr$\th]}

In the previous subsections we assumed that we can see the prompt emission of
GRBs only if we are within the jet opening angle, \thv$\le$\th, 
i.e. that off--axis observers see no emission.  
This implies large bulk Lorenz factors, i.e. $\Gamma\gg 1/\theta_{\rm j}$. 
However, the slope of the LF found by WP10 below the break (dashed line in 
Fig. \ref{fg1})
is $\sim$1.25. 
This slope can be predicted in the case of UJ observed off--axis. 
To see this, consider a population of bursts $N(L_{\rm iso, on})$
that have the same $L_{\rm iso, on}$ when observed on axis.
The number of bursts seen at different viewing angles will depend on
the corresponding solid angle, and they will be observed with a reduced
(de--beamed) luminosity $L$
(see Ghisellini et al 2006 for GRBs; Urry \& Shafer 1984 and Celotti et al. 1993 for blazars):
\begin{equation}
N(L, L_{\rm iso, on}) dL \, =\, {d\Omega \over 2\pi}\,  \to
N(L, L_{\rm iso, on} ) \, =\, \left( {dL \over d\cos\theta} \right)^{-1}
\label{nloff}
\end{equation}

The observed luminosity is the sum of the 
contributions of different portions of the emitting surface, each observed under a different
angle (Ghisellini et al. 2006): 
\begin{equation}
L(\theta_{\rm v}) = \int_{\rm max(0,\theta_{\rm v}-\theta_{\rm j})}^{\theta_{\rm v}+
\theta_{\rm j}} \Delta \phi\, \delta^4\, L^\prime\, \sin\theta\,  d\theta
\label{onoff}
\end{equation}
where $L^\prime$ is the comoving luminosity per unit solid angle
(which is assumed constant throughout the entire jet surface), 
$\theta$  is defined as the angle between the line of sight and each  
emitting element of the jet and $\Delta \phi$ takes into account the  
geometry of the emitting surface. 
If the line of sight coincides with the jet axis the emitting surface is a circular corona so that 
$\Delta \phi=2\pi$, while for off--axis observers this is a portion of circular corona 
which depends on \thv\ (Ghisellini \& Lazzati 1999):
\begin{equation}
\Delta\phi =
\begin{cases}
2\pi  & \text{if $\theta < \theta_{\rm j}-\theta_{\rm v}$} \\
\pi + 2\sin \left(\frac{\theta_{\rm j}^{2}-\theta_{\rm v}^{2}-\theta^2}
{2\theta_{\rm v}\, \theta}\right)  & \text{if $\theta \ge \theta_{\rm j}-\theta_{\rm v}$} \\
\end{cases}
\label{deltaphi}
\end{equation}

The right panel of Fig. \ref{fg4} shows the LF for different choices of \th\ and $\Gamma$. 
According to the different scaling of $L$ with $\theta_{\rm v}$ we have different 
slopes of the LF, which can be easily explained:

\vskip 0.2 cm
\noindent
{\it Small angles: $\theta_{\rm v}<\theta_{\rm j}$ ---}
 the observed luminosity increases
only slightly by decreasing $\theta_{\rm v}$.
This describes the high luminosity end of the LF, in which $N(L)$ corresponds
to the GRBs observable within $\theta_{\rm jet}$ (see the right panel of Fig. \ref{fg4}).

\vskip 0.2 cm
\noindent
{\it Intermediate angles: $\theta_{\rm j}<\theta_{\rm v}\lsim\ 2\theta_{\rm j}$ ---}
Eq. \ref{onoff} shows that the luminosity depends on the bulk Lorentz factor $\Gamma$, 
on the jet aperture angle \th, and $L^\prime$. Assuming $L^\prime=10^{49}$ erg s$^{-1}$, 
Fig. \ref{fg4} (left panel)
shows how $L(\theta_{\rm v})$ depends on the different choices of \th\ and  $\Gamma$.
For large $\Gamma$ the observed luminosity drops by a large factor when
$\theta_{\rm v}$ becomes slightly larger than $\theta_{\rm j}$. 
This drop is smoother for smaller $\Gamma$.
For $\theta_{\rm j}<\theta_{\rm v}\lsim\ 2\theta_{\rm j}$ (namely, within the ``jump",
and before reaching the regime $L\propto \delta^4$), we can approximate the scaling of 
the luminosity as $L\propto \theta_{\rm v}^{-f}$.
According to Eq. \ref{nloff}, we have:
\begin{eqnarray}
N(L) &\propto&  \left[ {dL \over \sin\theta d\theta} \right]^{-1} \propto \theta^{(f+2)} 
\nonumber \\
      &\propto& L^{-(1+2/f)},\quad \theta_{\rm j}< \theta\ \lsim\ 2\theta_{\rm j}
\end{eqnarray}
Note that the slope $1+2/f$ tends to unity for large $f$.

\vskip 0.2 cm
\noindent
{\it Large angles: $\theta_{\rm v}\gg \theta_{\rm j}$ ---}
When $\theta_{\rm v}\gg \theta_{\rm j}$, the burst can be considered to 
have a mono--directional velocity, along the jet axis.
In this case the observed luminosity is proportional to $\delta^4$,
where $\delta \equiv [\Gamma(1-\beta\cos\theta_{\rm v})]^{-1}$ is the 
relativistic Doppler factor.
Therefore $(dL/d\cos\theta)^{-1}\propto \delta^{-5} \propto L^{-5/4}$ 
[this results is general, as demonstrated by Urry \& Shaefer
(1984): for beaming amplification factors $L\propto \delta^p$, the
resulting low luminosity branch of the observed luminosity function
$N(L) \propto L^{-(1+1/p)}$].

As a first approximation consider unique values of $\Gamma$ and $\theta_{\rm j}$ for all GRBs.
The shapes of the corresponding LF are shown in Fig. \ref{fg4} for different combinations of these two parameters, assuming a value of $L'=10^{49}$ erg s$^{-1}$. The LF is a power--law with a slope $-5/4$ which smoothly turns into a peak at high luminosities corresponding to the maximum observable luminosity. The LF of GRBs shown in Fig.\ref{fg1} is consistent with a power--law of slope $-1.2$ (WP10) and breaks into a steeper ($-2.4$) power--law for $L_{\rm iso} > 10^{52.5}$ erg s$^{-1}$. Therefore, there is a very good agreement between the theoretical LF slope (Fig. \ref{fg4}) and the observed one below $10^{52.5}$ erg s$^{-1}$ (Fig.\ref{fg1}). If we assume that the break luminosity corresponds to $L_{\rm iso,on}$, i.e. of GRBs seen within the cone of their jet (i.e. those making the peak of the LF curves in Fig.\ref{fg4}), the true rate of GRBs can be computed as $N_{\rm tot} = N_{\rm obs}/(1-\cos\theta_{\rm j})$ (where $N_{\rm obs}$ is the rate corresponding to the break of the LF of Fig. \ref{fg1}).  Assuming that only $\sim$ 0.3\% of SNIb/c produce GRBs (e.g. Ghirlanda et al. 2013), we estimate $\theta_{\rm j} \sim 4^\circ$. The other two parameters $\Gamma$ and $L'$ of the model can be constrained requiring that the integral of the LF corresponds to $N_{\rm tot}\sim100$ GRBs yr$^{-1}$ Gpc$^{-3}$. 

\begin{figure}
\hskip -1truecm
\psfig{file=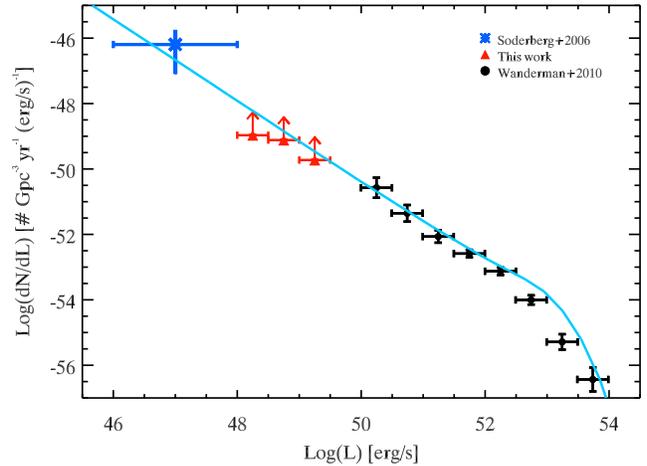,width=9.8cm}
\vskip -0.3 cm
\caption{Uniform jet observed in and out (ref \S3.3). }
\label{fg5}
\end{figure}

Still the model is not representative of the real data points, since there is the peak of the LF at the maximum luminosity (i.e. corresponding to the bursts observed within the jet aperture angle). In order to smooth this peak and obtain a break in the LF model, we introduce some dispersions of the values of $\Gamma$ and $\theta_{\rm j}$. We find a good agreement with the data as shown in Fig. \ref{fg5} by the solid line assuming $\theta_{\rm j}$ centered around 3$^\circ$ with a log--normal dispersion of width 0.2 and $\Gamma$ centered at a value of 30 with a log--normal dispersion of width 0.14.

\begin{figure}
\hskip -0.7truecm
\psfig{file=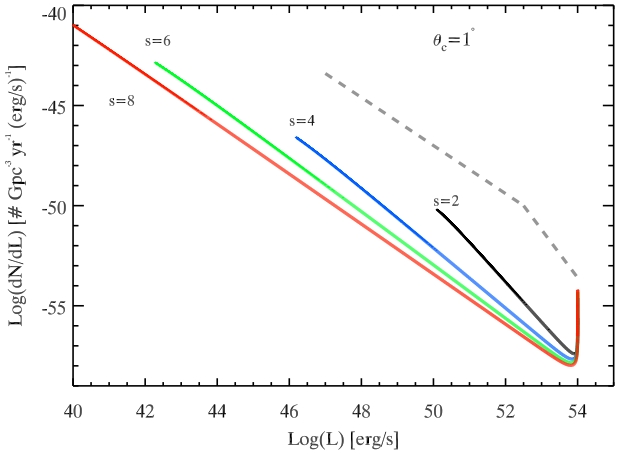,width=9.5cm}
\vskip -0.3 cm
\caption{Structured Jet model (solid lines) for different values of the power law energy structure slope $k$ (as labeled) assuming a core angle of 1$^\circ$. The dashed grey line shows the observed luminosity function (arbitrarily rescaled for clarity) of WP10 extending from LL to HL.}
\label{fg6}
\end{figure}

\begin{figure}
\hskip -1.2truecm
\psfig{file=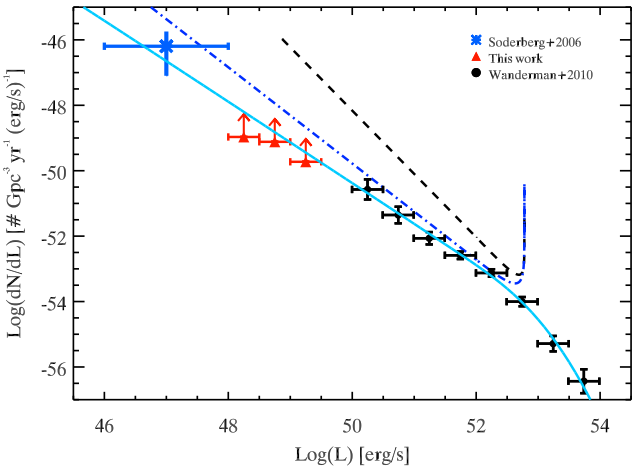,width=9.8cm}
\vskip -0.3 cm
\caption{Structured Jet model assuming a jet core energy per unit solid angle $\epsilon_{\rm c}=6\times10^{52}$ erg with a dispersion (log--normal) around this value with $\sigma$=0.5, $s=8.1$ and $\theta_{\rm c}\sim5^\circ$. For comparison are also shown the LF obtained with $s=2$ and $s=4$ (from Fig.\ref{fg6}) by the dashed and dot--dahes lines, respectively.}
\label{fg7}
\end{figure}

\section{Structured jet}

In this section we investigate the case of a structured jet (SJ),
assuming that all burst have an identical structure, and that 
 their observed properties 
are only due to the viewing angle \thv. 
We assume that the distribution of the energy per unit solid 
angle $\epsilon(\theta)$ is constant within a core of small aperture $\theta_c$ 
and distributed as a power law at larger angles:
\begin{equation}
\epsilon(\theta) =
\begin{cases}
\epsilon_{c}  & \text{if $\theta \leq \theta_{c}$} \\
\epsilon_{c}\left(\frac{\theta}{\theta_{c}}\right)^{-s}   & \text{if $\theta > \theta_{c}$}
\end{cases}
\label{eq4}
\end{equation}
where $\epsilon_{c}$ is the energy per unit solid angle within the core of the jet. 
The integral of $\epsilon(\theta_v)$ is the total jet energy $E$ which is a free parameter. 
Different jet structures have been considered from $s=2$ (Rossi et al. 2002) 
to steeper structures (Zhang \& Meszaros 2002 considered also the limit of a gaussian jet). 
In the following $s$ is a free parameter. 
We define the energy observed at a certain \thv\ as 
$E_{\rm iso}(\theta_{\rm v})=4\pi\eta\epsilon(\theta_{\rm v})$ and assuming a typical 
duration we can define the luminosity as 
$L_{\rm iso}(\theta_{\rm v}) \propto \epsilon(\theta_{\rm v})$. 
For $\theta_{\rm v}\le \theta_{c}$, this corresponds to the maximum luminosity 
$L_{\rm iso,max}\propto \epsilon_{c}$. 
The minimum luminosity  is for  $\theta_{\rm v}=\pi/2$. 
From Eq. \ref{eq4} the ratio between the maximum and minimum luminosity is 
$L_{\rm max}/L_{\rm min} = (\pi/2\theta_{c})^s$: 
the extension of the resulting LF is set by the slope $s$ and the jet core $\theta_c$. 
Since the observed LF of GRBs (including HL and LL bursts) extends over seven orders of 
magnitudes (from $10^{47}$ erg s$^{-1}$ to $10^{54}$ erg s$^{-1}$), we derive the limit
$s \gtrsim 7/\log [\pi/(2\theta_c)]$. 
A typical core angle $\theta_c=1^\circ$ implies $s>3.6$, thus excluding the $s=2$ 
case which is instead what required by the clustering of the collimation corrected 
energies of GRBs (Frail et al. 2002). 

The luminosity function of the SJ results:
\begin{eqnarray}
P(L_{\rm iso}) &\propto& 
\sin\left\lbrace \left( \frac{L_{\rm iso}}{L_{\rm iso,max}} \right)^{-\frac{1}{s}} 
\theta_{c} \right\rbrace \left( \frac{L_{\rm iso}}{L_{\rm iso,max}} \right)^{-1-\frac{1}{s}} 
\nonumber \\
 &\approx &
\theta_{c} \left( \frac{L_{\rm iso}}{L_{\rm iso,max}} \right)^{-(1+2/s)} 
\label{eq8}
\end{eqnarray}
since the argument of the sine is always small for reasonable values of $\theta_{\rm c}$.
The of the LF depends on the shape of the jet structure $s$ which also regulates 
the extension of the LF (i.e. the ratio of the minimum and maximum observable luminosity) 
as shown by the different curves of Fig. \ref{fg6}. 
Formally, a slope $a=1.25$ of the observed LF of Fig. \ref{fg1} we get a typical value of $s=8$. 
The upturn at large luminosities shown in Fig. \ref{fg6} corresponds to the jets observed within 
the core which all have the same luminosity. 
Instead, the observed LF (Fig. \ref{fg1}) is steeper after a break corresponding 
to $L_{\rm iso}\sim 3\times10^{52}$ erg s$^{-1}$. 
To reproduce this smooth break we introduce some dispersion of the parameters. 
Fig. \ref{fg7} shows that we can reproduce the LF  if we assume a jet core energy per unit solid angle $\epsilon_{\rm c}=6\times10^{52}$ erg with a dispersion (log--normal) around this value with $\sigma$=0.5, $s=8.1$ and $\theta_{\rm c}\sim5^\circ$. The obtained LF is also consistent with the lower limits corresponding to the IL bursts.

\begin{table*}
\begin{center}
\begin{tabular}{llllllllllllll}
\hline\hline
Model   &Fig.  &Main requirement  &Good fit? &Comment \\
\hline
Uniform, same $\theta_{\rm j}$         &\ref{fg2} &Flat below $\eta L_\star/\theta_{\rm j}^2$        &no  &No reason. param. values \\
Uniform, $\theta_{\rm j}(L_{\rm iso})$ &\ref{fg3} &Needs $\theta_{\rm j}\propto L_{\rm iso}^{-\xi}$, $\xi=0.5$ or 0.3 &yes &Marg. consistent with LL     \\
Uniform, on/off--axis                  &\ref{fg5} &Needs distrib. of $\theta_{\rm j}$ and $\Gamma$  &yes &Marg. consistent with HL \\
Structured                             &\ref{fg7} &Needs $\epsilon \propto \theta^{-4}$ or steeper  &yes &$\epsilon \propto \theta^{-2}$ excluded      \\
\hline\hline
\end{tabular}
\end{center}
\caption{
Synoptic view of our findings.
}
\label{tab2}
\end{table*}%

\section{Discussion and Conclusions}

Due to the increased capability to measure the redshift of long GBRs, we have 
now better determination of their luminosity function, and indeed the
results of different groups and of different methods start to converge.
The LF can be modeled as a broken power law, with slopes $\sim$1.2--1.5 and
$b>2$, and a break above $10^{52}$ erg s$^{-1}$.
The degree of cosmic evolution is instead still uncertain (see e.g. S12).
Another controversial issue concerns low luminosity GRBs.
We have observed very few of them, but their vicinity points
to a very large rate.
From what we find, they lie on the extrapolation of the luminosity
function that describes high luminosity events.
This suggests that LL and HL bursts belong to the same population.
LL bursts have luminosities of $\sim10^{47}$ erg s$^{-1}$ and
therefore extend the range of observed GRB luminosities to
seven orders of magnitude.
Is this very large range produced by the different intrinsic GRB power 
or is it instead the result of viewing the
same intrinsic phenomenon under different lines of sight?
We here discuss our findings, that are summarized
in Tab. \ref{tab2}.

\begin{itemize}
\item
If the jets of GRBs are homogeneous and have a typical opening angle,
then the observed luminosity is proportional to the 
energetics of the bursts after it has spent part of its
initial energy to punch the progenitor star.
This implies that the LF must be flat at low luminosities,
and this contradicts the data.
Our conclusion is that we can exclude this simple case.

\item
We have then examined the case of a homogeneous jet, but
with a jet angle that is related to the GRB energetics:
smaller $\theta_{\rm j}$ correspond to larger $E_{\rm iso}$ and
$L_{\rm iso}$. 
Since low luminosity bursts have larger $\theta_{\rm j}$ 
the probability that they intercept our line of sight is greater then
that for high luminosity bursts:
therefore the fraction of GRBs that we detect at lower luminosities is
greater than at large luminosities.
If the ``true" LF is flat, 
the observed LF is instead decreasing towards larger $L_{\rm iso}$ with a
slope that depends on the chosen relation between $\theta_{\rm j}$ 
and $L_{\rm iso}$.
The latter can be inferred by the observed spectral energy correlations,
that yields $\xi=0.5$ (in the case of a circumburst wind density profile)
or $\xi=0.3$ (homogeneous circumburst density). 
In this case we can obtain a reasonable agreement with the data
in the entire luminosity range but the very low luminosities,
where the model shows a small deficit.

\item
In the third case the jet is still homogeneous, 
but it can be seen also off axis, for viewing angles $\theta \gsim\theta_{\rm j}$.
This is possible for $\Gamma$ not extreme, as illustrated in Fig. \ref{fg4}:
the luminosity jump at $\theta_{\rm v}\gsim \theta_{\rm j}$ is more pronounced
for larger $\Gamma$.
The LF constructed with a single value of $\theta_{\rm j}$ and $\Gamma$
do not fit the data, since they show an upturn at high luminosity instead
of a steepening. 
On the other hand, assuming some dispersion of $\theta_{\rm j}$ and $\Gamma$,
we can obtain a reasonable agreement.
The required average values are $<\Gamma> =30$ and $<\theta_{\rm j}>= 3^\circ$.
What is remarkable is that the analytically predicted slope
at intermediate luminosities is very close to what seen.

\item
Finally, we have investigated structured jets,
of the form $\epsilon(\theta_{\rm v}) \propto \theta_{\rm v}^{-s}$
beyond a core angle $\theta_{\rm c}$.
We have found that a good fit can be obtained, but only if the slope
$s$ is rather steep, $s>4$, with a preferred value $s\sim 8$.
This is much steeper than the value $s=2$ originally proposed
to explain the clustering of $E_\gamma$ found by Frail et al. (2001).

\end{itemize}

These studies indicate that the jet must have a relatively sharp cut--off. 
Even if an abrupt one is unphysical (all the energy contained within $\theta_{\rm j}$, and
zero outside), the energy must in any case decrease rapidly with the angle
from the jet axis, once it it becomes greater than the core angle $\theta_{\rm c}$.
The other important overall conclusion is that although the low luminosity 
bursts seem not to have enough energy to punch the progenitor star, they 
can nevertheless understood within the same framework of large luminosity
GRBs, as long as they have a larger jet angle, or they are seen off--axis.
In the first case we see the little energy leftover after the jet break--out, 
in the latter case the apparent low luminosity is due to the large viewing angle, 
but the real energetics of these burst in much greater.


\section*{Acknowledgments}
We acknowledge ASI I/004/11/0 and the 2011 PRIN-INAF grant for
financial support.

\end{document}